\documentclass[conference,10pt]{IEEEtran}
\IEEEoverridecommandlockouts
\usepackage{amsmath,amssymb,amsfonts}
\usepackage[shortlabels]{enumitem}
\usepackage[lined,ruled,linesnumbered]{algorithm2e}
\SetKwComment{Comment}{/* }{ */}
\usepackage{algorithmic}
\usepackage{graphicx}
\usepackage{textcomp}
\usepackage{xcolor}
\usepackage{bm}
\usepackage{epsfig}
\usepackage{subfigure}
\usepackage{psfrag}
\usepackage{cite}
\usepackage[normalem]{ulem}

\def\BibTeX{{\rm B\kern-.05em{\sc i\kern-.025em b}\kern-.08em
		T\kern-.1667em\lower.7ex\hbox{E}\kern-.125emX}}

\begin{document}
\title{{\color{black} NCAirFL: CSI-Free Over-the-Air Federated Learning Based on Non-Coherent Detection}}
\author{\IEEEauthorblockN{Haifeng Wen${}^{\ast}$, Nicolò Michelusi${}^{\ddagger}$,  Osvaldo Simeone${}^\dagger$, and Hong Xing${}^{\ast\S}$
}\\
	\IEEEauthorblockA{${}^\ast$ The Hong Kong University of Science and Technology (Guangzhou), Guangzhou, China \\
         ${}^{\ddagger}$ Arizona State University, Arizona, U.S. \\ 
         ${}^\dagger$ KCLIP Lab, Centre for Intelligent Information Processing Systems (CIIPS),\\ King's College London, London, U.K.\\
         ${}^\S$ The Hong Kong University of Science and Technology, HK SAR, China \\
		E-mails:~hwen904@connect.hkust-gz.edu.cn,~nicolo.michelusi@asu.edu,~osvaldo.simeone@kcl.ac.uk,~hongxing@ust.hk
  }
}

\maketitle

\begin{abstract}
Over-the-air federated learning (FL), i.e., AirFL, leverages computing primitively over multiple access channels. A long-standing challenge in AirFL is to achieve coherent \emph{signal alignment} without relying on expensive channel estimation and feedback. This paper proposes \emph{NCAirFL}, a CSI-free AirFL scheme based on unbiased non-coherent detection at the edge server. By exploiting \emph{binary dithering} and a long-term memory based error-compensation mechanism, NCAirFL achieves a convergence rate of order $\mathcal{O}(1/\sqrt{T})$ in terms of the average square norm of the gradient for general non-convex and smooth objectives, where $T$ is the number of communication rounds. Experiments demonstrate the competitive performance of NCAirFL compared to vanilla FL with ideal communications and to coherent transmission-based benchmarks.
\end{abstract}


\IEEEpeerreviewmaketitle

\newtheorem{definition}{\underline{Definition}}[section]
\newtheorem{fact}{Fact}
\newtheorem{assumption}{Assumption}
\newtheorem{theorem}{\underline{Theorem}\underline{\hspace{1.5em}}\hspace{-1.5em}}[section]
\newtheorem{lemma}{\underline{Lemma}}[section]
\newtheorem{proposition}{\underline{Proposition}}[section]
\newtheorem{corollary}[proposition]{\underline{Corollary}}
\newtheorem{example}{\underline{Example}}[section]
\newtheorem{remark}{\underline{Remark}}[section]
\newcommand{\mv}[1]{\boldsymbol{#1}}
\newcommand{\mb}[1]{\mathbb{#1}}
\newcommand{\Myfrac}[2]{\ensuremath{#1\mathord{\left/\right.\kern-\nulldelimiterspace}#2}}
\newcommand\Perms[2]{\tensor[^{#2}]P{_{#1}}}

\section{Introduction} \label{sec:Introduction} 
Federated learning (FL) can support decentralized AI implementations across siloed data centers or across distributed IoT terminals \cite{mcmahan2017communication, chen2023knowledge, chen2024robust}. \emph{Over-the-air FL (AirFL)} has emerged as a promising integrated communication and computation technique that enables communication-efficient model aggregation by exploiting the waveform superposition property of wireless channels \cite{nazer2007computation, Bingnan2024overtheair}.
However, AirFL relies on the alignment of signals from different devices.
Truncated channel inversion, or phase compensated equalization, can be applied at the transmitter side to ensure the alignment of the received signals, but at the cost of receiving channel state information (CSI) at the transmitter (CSIT)~\cite{zhu2019broadband,sery2020analog,Ni2022STAR}.

Channel inversion-based transmissions for AirFL suffer from several drawbacks:
(\emph{i}) additional pilot overhead and feedback are required to obtain CSIT, with requirements linearly increasing with the number of devices and AI model dimensions \cite{ma2003optimal, zhu2019broadband};  
and (\emph{ii}) the power of the effective noise is amplified due to deep fades \cite{zhu2019broadband}. 
While phase-compensated equalization can avoid issue (\emph{ii}), it was only studied for narrow-band transmissions and may not be directly appliable to AirFL over broadband channels \cite{sery2020analog}.
Other AirFL transmission protocols entailing CSIT include precoding optimization \cite{yang2020federated} and device selection \cite{xia2021fast}.

To address the problem (\emph{i}), CSIT-free transmissions were studied in~\cite{amiri2021blind, yang2021revisiting, wei2023random, choi2022communication}. 
References \cite{amiri2021blind} and \cite{yang2021revisiting} investigated the effect of channel fading on the learning performance of CSI-free AirFL training, revealing that CSI-free transmissions induce an error floor in the convergence bounds.
The authors of \cite{wei2023random} and \cite{choi2022communication} proposed to alleviate the burden of channel estimation by leveraging massive multiple-input and multiple-output (MIMO) techniques via channel hardening.

Recently, a CSI-free over-the-air implementation of decentralized stochastic gradient descent (DSGD) scheme was proposed in \cite{michelusi2024non} for fully decentralized FL with only wireless device-to-device (D2D) link among terminals. The proposed scheme enjoys a convergence rate of $\mathcal{O}(1/\sqrt{T})$ after $T$ communication rounds for \emph{strongly-convex} training objectives leveraging cross-polytope-$\phi$ codebook and unbiased non-coherent detection.
However, the transmission overhead for its codewords was shown to be around two times as large as the model dimension to achieve good performance, thus compromising communication efficiency. 

Inspired by \cite{michelusi2024non}, in this paper, we propose a CSI-free implementation of AirFL referred to as \emph{NCAirFL}, which adopts a novel \emph{binary dithering} coding together with unbiased non-coherent detection and a long-term error-feedback mechanism.
The convergence rate of the proposed NCAirFL for general $L$-smooth non-convex losses is shown to be $\mathcal{O}(1/\sqrt{T})$, coinciding with FedAvg under ideal communications~\cite{mcmahan2017communication}.
Finally, numerical results demonstrate that NCAirFL obtains performance superior to other benchmarks based on coherent transmissions.  


\section{System Model} \label{sec:System Model} 
In this section, we consider an AirFL system consisting of a set $[n]\triangleq\{1, \ldots, n\}$ of devices and an edge server over a broadband Gaussian fading channel. First, we provide preliminaries on vanilla \emph{FedAvg} with ideal, noiseless, communications, and then introduce the non-coherent detection-based communication model.

\subsection{Learning Protocol (FedAvg)} \label{subsec:Learning Model}
The $n$ devices collaboratively train a shared machine learning model (e.g., a deep neural network or a large language model) parameterized by $\mv \theta$  to solve an empirical loss minimization problem given by
\begin{align*}
\mathrm{(P0)}:&~\mathop{\mathtt{Minimize}}_{\mv \theta \in \mathbb{R}^d}~~~f(\mv \theta)\triangleq\frac{1}{n}\sum_{i=1}^{n}f_{i}(\mv \theta).
\end{align*} 
In $\mathrm{(P0)}$, $f(\mv \theta)$ represents the global empirical loss function, while $f_{i}(\boldsymbol \theta) = 1/\vert\mathcal{D}_{i}\vert\sum_{\boldsymbol\xi\in\mathcal{D}_{i}}\ell (\boldsymbol \theta;\boldsymbol\xi)$ represents the local empirical loss function at device $i \in [n]$; 
\textcolor{black}{$\mathcal{D}_{i}$ denotes the local private data set with size $\vert\mathcal{D}_{i}\vert$ at device $i\in [n]$;}
and the loss function $\ell (\boldsymbol{\theta};\boldsymbol{\xi})$ indicates the loss accrued from parameter $\boldsymbol{\theta}$ with respect to (w.r.t) the data sample $\boldsymbol{\xi}$.

In the standard FedAvg protocol~\cite{mcmahan2017communication}, each device first updates the individual model $\mv \theta_i$, $i\in[n]$, using the local data set $\mathcal{D}_i$, and then the centralized server averages over the model updates to obtain the next iterate $\mv \theta$. The above procedure is repeated for $T$ iterations or communication rounds, to obtain an approximate solution to problem $\mathrm{(P0)}$. 

Specifically, at the $t$-th communication round, $t\in\{0,1,\ldots,T-1\}$, the server selects a fraction $r\in(0,1]$ of devices uniformly at random, forming the subset $\mathcal{I}^{(t)}$ of active devices with $|\mathcal{I}^{(t)}|=rn$.
Each active device in $\mathcal{I}^{(t)}$ executes $Q$ local SGD steps over a mini-batch $\mathcal{B}_i^{(t,q)}\subseteq \mathcal{D}_i$
of data samples commencing with shared initialization $\mv \theta_i^{(t,0)}\leftarrow\mv \theta^{(t)}$ by the server, i.e.,
\begin{equation} \label{eq:local updates}
    \boldsymbol \theta_{i}^{(t, q+1)} \leftarrow \boldsymbol \theta_{i}^{(t, q)}-\eta^{(t)}\hat\nabla f_{i}(\boldsymbol \theta_{i}^{(t, q)}),
\end{equation}
where $q\in\{0,\ldots, Q-1\}$ is the local iteration index; $\eta^{(t)}>0$ denotes the learning rate; and $\hat\nabla f_{i}(\boldsymbol \theta_{i}^{(t,q)})$ is the estimate of the true gradient $\nabla f_{i}(\boldsymbol \theta_{i}^{(t,q)})$ given by
$$
\hat\nabla f_{i}(\boldsymbol \theta_{i}^{(t, q)})=\frac{1}{\vert\mathcal{B}_{i}^{(t,q)}\vert}\sum\limits_{\boldsymbol\xi\in\mathcal{B}_{i}^{(t,q)}}\nabla \ell (\boldsymbol \theta_{i}^{(t, q)};\boldsymbol\xi).
$$

After local updating, each device $i\in\mathcal{I}^{(t)}$ transmits the model difference
\begin{equation}\label{eq:modeldiff}
\mv{\Delta}_{i}^{(t)} = \mv \theta_{i}^{(t,0)}-\mv \theta_{i}^{(t, Q)}
\end{equation} 
to the server. Subsequently, the server averages over the differences \eqref{eq:modeldiff} from  the active device set to update the global model parameter as
\begin{equation}\label{eq:global updates}
    \mv \theta^{(t+1)} \leftarrow \mv \theta^{(t)} - \frac{1}{rn}\sum_{i \in \mathcal{I}^{(t)}}\mv{\Delta}_{i}^{(t)}.
\end{equation} 

\subsection{Communication Model} \label{subsec:Communication Model}
At the $t$-th communication round, all devices in $\mathcal{I}^{(t)}$ simultaneously transmit their local update $\mv \Delta_i^{(t)}$ to the edge server over the air using orthogonal frequency division multiplexing (OFDM).
To elaborate, define as $\mathcal{C}:\mathbb{R}^d \mapsto \mathbb{R}_{+}^d$ a preprocessing function that maps the devices' local update $\mv \Delta_i^{(t)}$ to a non-negative vector $\mv g_i^{(t)}$, $i\in \mathcal{I}^{(t)}$. The design of the preprocessing function $\mathcal{C}(\cdot)$ will be discussed in the next section.
Each entry $j\in\{1, \ldots, d\}\triangleq [d]$ of the transmitted signal $\mv x_i^{(t)} = [ x_{i,1}^{(t)}, \ldots, x_{i,d}^{(t)}]^T$ for $i\in\mathcal{I}^{(t)}$ is given by
\begin{equation} \label{eq:non-coherent transmitted signal}
    x_{i,j}^{(t)} = \textcolor{black}{\alpha_i^{(t)}}\sqrt{\Myfrac{g_{i,j}^{(t)}}{\eta^{(t)}}},
\end{equation}
where $\textcolor{black}{\alpha_i^{(t)}}$ is a power \textcolor{black}{scaling} factor, and $g_{i,j}^{(t)}$ is the $j$-th entry of vector $\mv g_i^{(t)}$.

Assuming that $d$ subcarriers are allocated for the transmissions of the active devices in set $\mathcal{I}^{(t)}$, the received signal is given by~\cite{zhu2019broadband}
\begin{equation} \label{eq:received signal}
    \boldsymbol y^{(t)}=\sum_{i\in\mathcal{I}^{(t)}} \sqrt{\kappa_i}\boldsymbol h_{i}^{(t)}\odot \boldsymbol x_{i}^{(t)} + \boldsymbol n^{(t)},
\end{equation}
where $\kappa_{i}$ denotes the large-scale fading between device $i$ and the edge server; \textcolor{black}{$\boldsymbol h_{i}^{(t)}=[h_{i,1}^{(t)}, \ldots, h_{i,d}^{(t)}]^T$ is the $d$-dimensional vector of small-scale fading channel coefficient of device $i$, which follows an arbitrary distribution with each entry $h_{i,j}^{(t)}$'s being zero mean and unit variance, i.e., \(\mathbb{E}[|h_{i,j}^{(t)}|^2]=1\); $\mv h_i^{(t)}$'s is also assumed to be independent of the other channel vectors $\mv h_j^{(t)}$ for $j \ne i$, and remains constant for the $t$-th communication round but may vary from one round to another;}
$\boldsymbol n^{(t)}$ is the additive Gaussian noise denoted by $\mv n^{(t)}\sim \mathcal{CN}(0, \sigma^2 \mv I_d )$; and $\mv x\odot \mv y$ denotes the Hadamard product, i.e., the entry-wise product, between two vectors $\mv x$ and $\mv y$.

Like (truncated) channel inversion for coherent transmission-based implementations of AirFL~\cite{zhu2019broadband}, the power scaling factor $\textcolor{black}{\alpha_i^{(t)}}$ in \eqref{eq:non-coherent transmitted signal} is set as
$
\textcolor{black}{\alpha_i^{(t)}}=\Myfrac{\sqrt{\rho^{(t)}}}{\sqrt{\kappa_i}},
$
$i \in \mathcal{I}^{(t)}$, where the scaling factor $\rho^{(t)}$ is selected to satisfy the average transmit power constraints
\begin{equation} \label{eq:power constraint}
    \frac{1}{d} \mathbb{E}\|\boldsymbol x_{i}^{(t)}\|^2 \le P_i.
\end{equation}
As a result, the $j$-th entry of the received signal in \eqref{eq:received signal} can be written as 
\begin{equation} \label{eq:simplified received signal}
    y_j^{(t)}= \sqrt{\rho^{(t)}} \sum_{i\in\mathcal{I}^{(t)}} h_{i,j}^{(t)} \sqrt{ g_{i,j}^{(t)} / \eta^{(t)}} + n_j^{(t)}.
\end{equation}

To detect the sum of the transmitted signals, $\sum_{i\in\mathcal{I}^{(t)}}\mv g_i^{(t)}$, without CSI information at the edge server, we adopt a \emph{square-law} detector \(|\cdot|^2\) at each of the $d$ subcarriers, and obtain the sufficient statistics
\begin{equation} \label{eq:NC receiver}
    r_j^{(t)} = \Myfrac{\left(\left|y_j^{(t)}\right|^2-\sigma^2\right)}{\rho^{(t)}},
\end{equation}
which provides an unbiased estimate of $\sum_{i\in \mathcal{I}^{(t)}} g^{(t)}_{i,j}/\eta^{(t)}$, $j\in [d]$. 

Finally, the server estimates the true sum of model differences  $\sum_{i\in\mathcal{I}^{(t)}} \mv \Delta_i^{(t)}$ via a decoder $\mathcal{D}:\mathbb{R}^d \mapsto \mathbb{R}^d$, yielding the estimate $\widehat{\mv \Delta}^{(t)}=\mathcal{D}(\mv r^{(t)})$. The decoder will be specified in the next section. The global model is finally updated as 
\begin{equation}
    \mv \theta^{(t+1)} \leftarrow \mv \theta^{(t)}-\frac{1}{rn}\widehat{\mv \Delta}^{(t)}.
    \label{eq:airfl global update}
\end{equation}

\section{Non-coherent AirFL (NCAirFL)} \label{sec:NCAirFL}

In this section, we propose a novel CSI-free communication-efficient AirFL scheme, referred to as \emph{NCAirFL}. 
The key challenge addressed by NCAirFL is the design of preprocessing and decoding mappings, $ \mathcal{C}(\cdot)$ and $ \mathcal{D}(\cdot)$, in the non-coherent transmission scheme presented in Sec.~\ref{subsec:Communication Model}.

NCAirFL is based on a form of multiplicative binary dithering whereby the preprocessing function $\mv g_i^{(t)}=\mathcal{C}(\mv \Delta_i^{(t)})$ is designed as 
\begin{equation} \label{eq:sparsified model different}
    g_{i,j}^{(t)} =\mathcal{C}(\Delta_{i,j}^{(t)}) = \left(\left(m_{i,j}^{(t)}+\Delta_{i,j}^{(t)}\right) \phi_j^{(t)}\right)^+, \ j\in[d],
\end{equation}
with $(x)^{+}=\max(0,x)$, where $m_{i,j}^{(t)}$ and $\Delta_{i,j}^{(t)}$ are the $j$-th entry of $\mv m_i^{(t)}$ and $\mv \Delta_i^{(t)}$, respectively, and vector $\mv \phi^{(t)} = [\phi_1^{(t)},\ldots,\phi_d^{(t)}]^T \in \{-1,1\}^d$ has i.i.d. entries 
\begin{equation} \label{eq:flipping vector}
    \phi_j^{(t)}=
    \begin{cases}
    1 & \text{w.p.}\ p, \\ 
    -1 & \text{w.p.}\ 1-p.
    \end{cases}
\end{equation}
The memory vector  $\mv m_i^{(t)} \in \mathbb{R}^d$ in \eqref{eq:sparsified model different} offers error compensation via the following update rule:
\begin{equation} \label{eq:memory update rule}
\boldsymbol m_i^{(t+1)}=
\begin{cases}
    \boldsymbol m_i^{(t)}+\boldsymbol \Delta_i^{(t)} - \boldsymbol \phi^{(t)}\odot\boldsymbol g_i^{(t)} & \text{if } i\in\mathcal{I}^{(t)}, \\ 
    \mv m_i^{(t)} & \text{otherwise}.
\end{cases}
\end{equation}

The random binary dithers are generated independently for each round $t$ by a pseudorandom generator whose seed is also available at the receiver. Using this information, the decoder $\widehat{\mv\Delta}^{(t)}=\mathcal{D}(\mv r^{(t)})$ removes the effect of the dither as
\begin{equation} \label{eq:approximate aggregated model different}
    \widehat{\Delta}_{j}^{(t)} = \eta^{(t)} \phi_j^{(t)} r_j^{(t)}, \ j \in [d],
\end{equation}
Substituting \eqref{eq:approximate aggregated model different} in \eqref{eq:airfl global update} leads to the following global update rule:
\begin{equation} \label{eq:NCAirFL global update}
    \boldsymbol \theta^{(t+1)} \leftarrow \boldsymbol \theta^{(t)} - \frac{\eta^{(t)}}{rn}\boldsymbol \phi^{(t)}\odot \boldsymbol{r}^{(t)}.
\end{equation}
The procedure of the proposed NCAirFL scheme is summarized in Algorithm~\ref{alg:Algorithm 1}.

Thanks to the dithers $\{\mv \phi^{(t)}\}$, the cascade of preprocessing and decoding mappings satisfies the following contraction property.
\begin{lemma}[Contraction] \label{lemma:contraction}
For any $i\in [n]$ and $t=0,1,\ldots, T-1$, the following inequality holds for the NCAIrFL update \eqref{eq:sparsified model different}-\eqref{eq:memory update rule}
\begin{multline}
\label{eq:contraction}
\mathbb{E}\left\| \boldsymbol m_i^{(t)}+\boldsymbol \Delta_i^{(t)} - \boldsymbol \phi^{(t)}\odot \mathcal{C}(\mv \Delta_i^{(t)}) \right\|^2 \\ \le  \textcolor{black}{\left(1-\lambda \right)}\| \boldsymbol m_i^{(t)}+\boldsymbol \Delta_i^{(t)}\|^2,
\end{multline} 
where the expectation is taken over the randomness of the dither $\boldsymbol \phi^{(t)}$, and  \textcolor{black}{$\lambda=\min(p, 1-p)$}. By minimizing the upper bound of \eqref{eq:contraction} on the probability $p$, the optimal choice of $p$ is easily seen to be $1/2$, which results in $\lambda = 1/2$.
\end{lemma} 

As we will prove in the next section, this property together with the unbiasedness property of $\mv r^{(t)}$ \eqref{eq:NC receiver} plays a 
crucial role in ensuring NCAirFL's convergence to a stationary point.

{\small
\begin{algorithm}[t] 
\SetKwInOut{Input}{Input}
\SetKwInOut{Output}{Output}
\SetKwBlock{DeviceParallel}{On devices $i \in \mathcal{I}^{(t)}$ (in parallel):}{end}
\SetKwBlock{localSGD}{for $q=0$ to $Q-1$ do}{end}
\SetKwBlock{OnServer}{On Server:}{end}
\caption{NCAirFL} \label{alg:Algorithm 1}
\textbf{Input:} learning rate $\eta^{(t)}$, power constraints $\{P_{i}\}_{i \in [n]}$, number of communication rounds $T$, number of local rounds $Q$, active device ratio $r$, and a pseudo-random random seed for generating dithering $\{\mv \phi^{(t)}\}$ \\
Initialize $\mv \theta_{i}^{(0)}=\mv \theta^{(0)}$ and $~\mv m_{i}^{(0)} = \mv 0$ for all $i \in [n]$ and $t=0$ \\
\While{$t < T$}{
Generate $\mv \phi^{(t)}$ via \eqref{eq:flipping vector}\;
\DeviceParallel{
$\mv{\theta}^{(t,0)}_{i}\leftarrow \mv{\theta}^{(t)}$\;
\localSGD{
$\mv{\theta}^{(t,q+1)}_{i}\leftarrow \mv{\theta}^{(t,q)}_{i} - \eta^{(t)} \hat\nabla f(\mv \theta_{i}^{(t,q)})$\;
}
$\mv{\Delta}_{i}^{(t)} = \mv \theta_{i}^{(t,0)}-\mv \theta_{i}^{(t, Q)}$\;
$g_{i,j}^{(t)} = ((m_{i,j}^{(t)}+\Delta_{i,j}^{(t)}) \phi_j^{(t)})^+$, for $j\in[d]$ \;
Update memory 
$$\mv m_i^{(t+1)}=\boldsymbol m_i^{(t)}+\boldsymbol \Delta_i^{(t)} - \boldsymbol \phi^{(t)}\odot\boldsymbol g_i^{(t)};$$ \\
Transmit $x_{i,j}^{(t)}=\sqrt{\Myfrac{\rho^{(t)} g_{i,j}^{(t)}}{(\kappa_i \eta^{(t)}})}$, for $j\in[d]$\;
}
\OnServer{
Receive 
$$
y_j^{(t)}= \sqrt{\rho^{(t)}} \sum_{i\in\mathcal{I}^{(t)}} h_{i,j}^{(t)} \sqrt{ g_{i,j}^{(t)} / \eta^{(t)}} + n_j^{(t)},
$$
for $j\in[d]$\;
Obtain $ r_j^{(t)} = \Myfrac{(|y_j^{(t)}|^2-\sigma^2)}{\rho^{(t)}}$, for $j\in[d]$\;
Global update
$$
\boldsymbol \theta^{(t+1)} \leftarrow \boldsymbol \theta^{(t)} - \frac{\eta^{(t)}}{rn}\boldsymbol \phi^{(t)}\odot \boldsymbol{r}^{(t)};
$$ \\
Broadcast $\mv \theta^{(t+1)}$ to all $n$ devices\;
}
$t \leftarrow t + 1$\;
}
\end{algorithm}
}

\section{Convergence analysis of NCAirFL} \label{sec:main results}
In this section, we study the convergence performance of the proposed NCAirFL protocol for non-convex and smooth loss functions.
To facilitate convergence analysis, we make the following standard assumptions that have been widely adopted in prior art \cite{basu19qsparse, yang2021achieving}.
\begin{assumption}[$L$-Smoothness] \label{assumption: L-smoothness}
    The local empirical loss function $f_{i}(\cdot)$, for $i\in[n]$, is differentiable and $L$-smooth, with $L>0$, i.e., $ \text{for all } \mv x, \mv y \in \mathbb{R}^d$, the loss $f_{i}(\cdot)$ satisfies
    \begin{equation}
        \left\|\nabla f_{i}(\mv x)-\nabla f_{i}(\mv y)\right\| \leq L\|\mv x-\mv y\|.
        \label{eq:L-smoothness}
    \end{equation}
\end{assumption}
\begin{assumption}[Bounded Variance and Second Moment] \label{assumption:bounded variance}
    For all $i\in[n]$, the stochastic gradient $\hat\nabla f_i(\mv\theta_i^{(t,q)})$ is unbiased with bounded variance and bounded average squared norm, i.e.,
\begin{equation}
    \begin{aligned}
    & \mathbb{E}\left[\left\|\hat\nabla f_{i}(\boldsymbol \theta_{i}^{(t, q)})-\nabla f_i(\boldsymbol \theta)\right\|^2\right] \leq \sigma_l^2,
    \end{aligned}
\end{equation}
    and
\begin{equation}
    \mathbb{E}\left[\left\|\hat\nabla f_{i}(\boldsymbol \theta_{i}^{(t, q)})\right\|^2\right] \leq G^2,
\end{equation}
\end{assumption}
where the expectation is taken over the randomness of mini-batch selection.

\begin{assumption}[Bounded Heterogeneity] \label{assumption:heterogeneity}
For any $\boldsymbol \theta\in \mathbb{R}^{d}$, the gradient of local empirical loss functions $f_i(\boldsymbol \theta)$, $i\in[n]$, satisfies the following inequality
\begin{equation}
    \frac{1}{n}\sum_{i=1}^{n}\left\|\nabla f_i(\boldsymbol \theta)-\nabla f(\boldsymbol \theta)\right\|^2 \leq \sigma_g^2,
\end{equation}
where $f(\mv \theta) = (\Myfrac{1}{n})\sum_{i=1}^{n}f_i(\mv \theta)$.
\end{assumption}

Next, we study convergence in terms of the expected gradient norm averaged over iterations, i.e., $\Myfrac{1}{T}\sum_{t=0}^{T-1} \mathbb{E}\| \nabla f(\boldsymbol\theta^{(t)}) \|^2$, based on the above assumptions~\cite{basu19qsparse}.

\begin{theorem}[Convergence] \label{theorem:convergence}
Suppose that Assumptions 1-3 hold and that the subset $\mathcal{I}^{(t)}$ of active devices is selected uniformly at random (without replacement) with size $|\mathcal{I}^{(t)}|=rn$. Let $\{\boldsymbol \theta^{(t)}\}$ be generated by NCAirFL for $T$ communication rounds with constant learning rate $\eta^{(t)}=\eta$ satisfying \textcolor{black}{$\eta \in(0, 1/(\sqrt{240}QL)]$}. Then, on average over the randomness of mini-batch sampling, dithering vectors, device selection, small-scale fading, and channel noise, NCAirFL satisfies the following inequality
\begin{multline} \label{eq:convergence bound}
\hspace{-0.15in}\frac{1}{T}\sum_{t=0}^{T-1}\mathbb{E}\|\nabla f(\boldsymbol \theta^{(t)})\|^2 \le \underbrace{\frac{8(f_0-f^*)}{T \eta Q}}_{\text{Initialization error}} +\underbrace{\frac{4\eta LG_e^2}{Qr^2n^2}}_{\text{Non-coherent detection error}}  \\ + \underbrace{4\eta L Q G^2 + 40\eta^2Q L^2 (\sigma_l^2+6Q\sigma_g^2)}_{\text{SGD error \& heterogeneity}}  \\
 + \underbrace{\frac{48 \eta^2 Q^2 L^2(1-\lambda^2)G^2}{r^2\lambda^2}}_{\text{Contraction error}},
\end{multline}
where $f_0=f(\mv \theta^{(0)})$, $f^*=\min_{\mv \theta}f(\mv \theta)$, $G_e^2=(rn)^2\tilde{G}^2 + \Myfrac{(4rn \tilde{G}^2)}{\mathrm{SNR}_{\min}} +\Myfrac{\tilde{G}^2}{(d \mathrm{SNR}_{\min}^2)}$
with $\tilde{G}^2$ and $\mathrm{SNR}_{\min}$ being given by$(8/\lambda^2-6)Q^2G^2$ and $\min_{i\in[n]}P_i\kappa_i/\sigma^2$, respectively.
\end{theorem}
\begin{IEEEproof}
    A sketch of the proof can be found in Appendix.
\end{IEEEproof}

The bound in (\ref{eq:convergence bound}) provides insights into the learning performance of NCAirFL. By accounting for the initialization error; the non-coherent detection error, which decreases with the average receive SNR, $\mathrm{SNR}_{\min}$, and $\lambda$; the error due to the randomness of the stochastic gradient and due to the heterogeneity of the local data sets, which increases with the variance $\sigma_l^2$ and/or data heterogeneity through $\sigma_g^2$; and the contraction error due to the compression implemented by the encoding and decoding mappings which depend on $\lambda$.
By setting the learning rate as $\eta=\mathcal{O}(1/\sqrt{T})$, the bound in \eqref{eq:convergence bound} indicates that NCAirFL converges to a stationary point at a rate of $\mathcal{O}(1/\sqrt{T})$, which is the same as that for FedAvg in the presence of non-convex $L$-smooth objectives \cite[Theorem 1]{yang2021achieving}.

\section{Experiments}\label{sec:Experiments}


In this section, we evaluate the performance of NCAirFL for settings involving $n = 20$ devices, aiming for showing its efficiency compared to well known AirFL schemes in the literature as benchmarks.
We use the standard MNIST data set for the task of classifying handwritten digits, which consists of $60000$ training samples and $10000$ test samples, with each being a $28\times28$ image.
Both i.i.d. and non-i.i.d. data distributions are considered. In the i.i.d. case, the local data set $\mathcal{D}_i$ is sampled uniformly at random (without replacement) from the MNIST training data set, $i\in[n]$. 
In contrast, in the non-i.i.d. case, the global data set is distributed such that each device only receives samples from two out of the ten classes of labeled data~\cite{mcmahan2017communication}. 
All devices train a common neural network model, which consists of an input layer with shape $(28,100)$, a fully connected layer with $100$ neurons with ReLU activation, and a softmax output layer, resulting in a total number $d=79510$ of training parameters. We adopt SGD as the optimizer. 

A simplified path loss model is adopted with  $\kappa_{i}=\Myfrac{c^2}{\left(4\pi f_c r_{i}\right)^2}$, $i\in[d]$, where $c$ denotes the speed of light, $f_c=2.4$\,GHz is the central frequency, and $r_{i}\sim\mathcal{U}(0, 100)$ is the distance between device $i$ and edge server. The small-scale fading channel $\mv h_i^{(t)}$ is assumed to follow Rayleigh distribution. 
Unless otherwise specified, the simulation parameters are set as follows: the mini-batch size is $\vert\mathcal{B}_{i}^{(t,q)}\vert=64$ for $q\in\{0, \ldots, Q-1\}$ and $i\in[n]$; the number of local iterations is $Q = 5$; the participation ratio is $r=0.2$ for i.i.d. case and $r=1$ for non-i.i.d. case; we set $p=1/2$ for the dithers; the power constraint is \(P_i = 2\times 10^{-8} \)\,W, for all $i \in [n]$; and the AWGN variance $\sigma^2=-123$\,dBm. 
We report the test-accuracy results averaged over 10 trials.

We consider the following three benchmarks: 
(\emph{i}) FedAvg with perfect communications \cite{mcmahan2017communication}; 
(\emph{ii}) the AirFL based on truncated channel inversion (truncated CI), which is referred to as \emph{CAirFL}~\cite{zhu2019broadband}; and (\emph{iii}) AirFL-Mem, a truncated-CI based AirFL with a long-term memory mechanism to compensate for the deep fading induced error~\cite{wen2024AirFL-Mem}. Table~\ref{table:convergence rate comparisons} summarizes the comparison of NCAirFL to these benchmarks in terms of convergence rate (for non-convex and smooth losses with constant learning rate $\eta=\mathcal{O}(1/\sqrt{T})$), error floor (indicating if a scheme can converge to a stationary point with arbitrarily small error),  and CSI requirements.
 
\begin{table*}[t] 
\centering
\caption{Porperties of the considered FL schemes} \label{table:convergence rate comparisons}
\begin{tabular}{llll} 
\hline
\textbf{Schemes}             & \textbf{Convergence Rate}         & \textbf{Error Floor} & \textcolor{black}{\textbf{CSI Requirements}} \\ \hline
FedAvg~\cite{mcmahan2017communication} & $\mathcal{O}(1/\sqrt{T})$         & No  & N/A                                                \\
CAirFL~\cite{zhu2019broadband}         & $\mathcal{O}(1/\sqrt{T})$         & Yes    & \textcolor{black}{CSIT}                           \\
AirFL-Mem~\cite{wen2024AirFL-Mem}                    & $\mathcal{O}(1/\sqrt{T})$         & No      & \textcolor{black}{CSIT}                           \\
\textbf{NCAirFL}             & $\mathcal{O}(1/\sqrt{T})$ & \textbf{No}     & \textcolor{black}{\textbf{No CSI}}                                \\ \hline
\end{tabular}
\end{table*}

\begin{figure}[t] 
    \centering
    \includegraphics[width=3.4in]{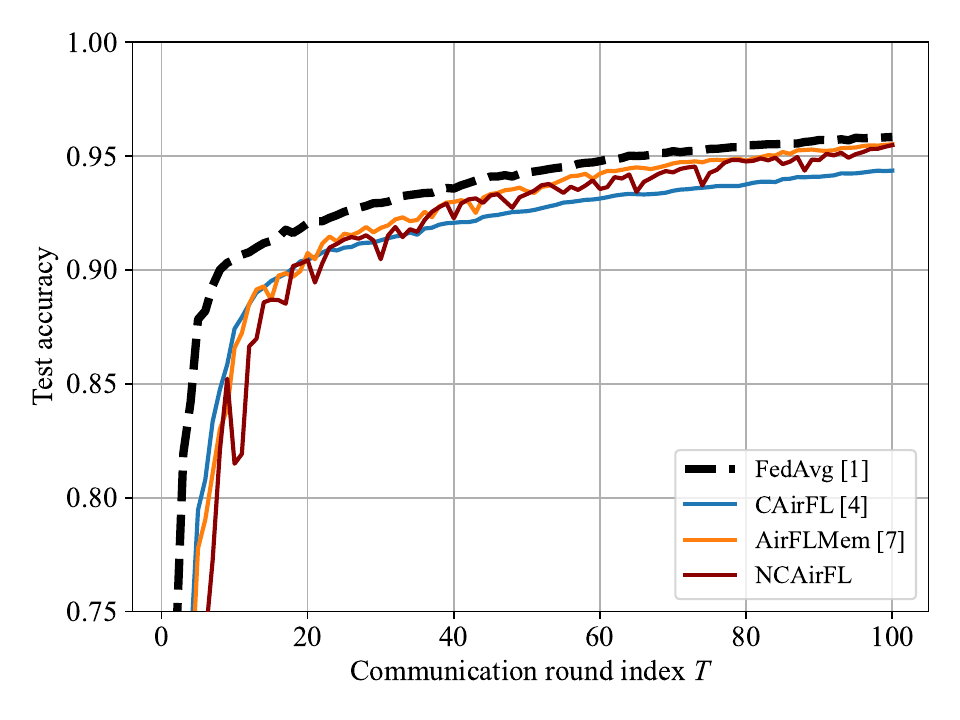}
    \caption{Test accuracy versus communication round index $T$ in the i.i.d. case.}
    \label{fig:acc_vs_T_iid}
\end{figure}

\begin{figure}[t] 
    \centering
    \includegraphics[width=3.4in]{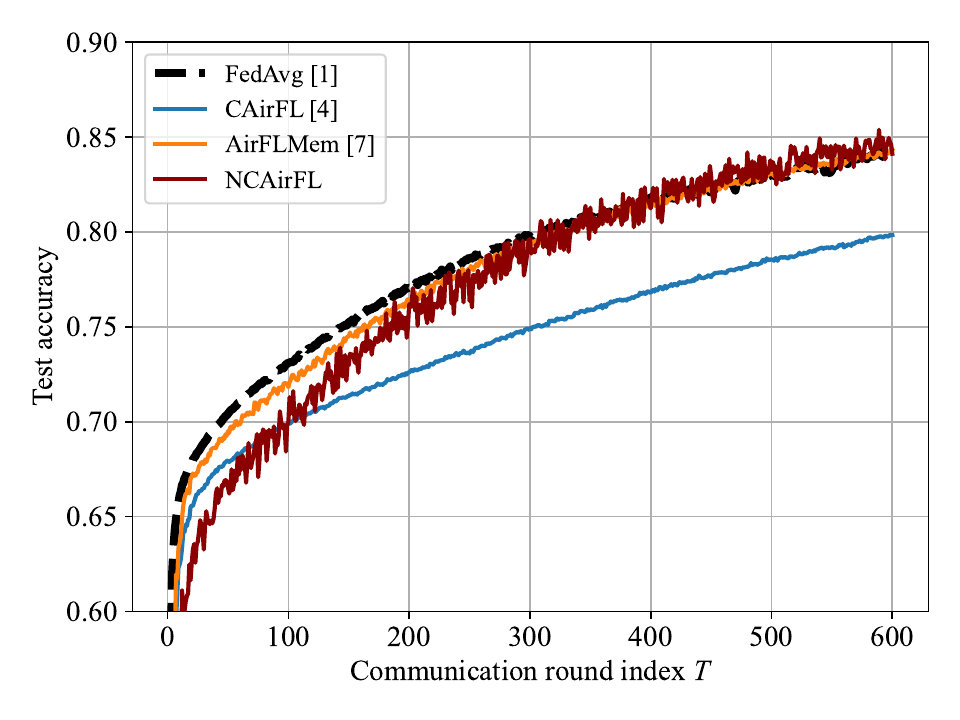}
    \caption{Test accuracy versus communication round index $T$ in the non-i.i.d. case.}
    \label{fig:acc_vs_T_pat}
\end{figure}

Fig.~\ref{fig:acc_vs_T_iid} and Fig.~\ref{fig:acc_vs_T_pat} show the test accuracy versus the communication round index $T$ for the proposed NCAirFL and for the benchmarks in the i.i.d. and the non-i.i.d. case, respectively. The results show that NCAirFL achieves performance comparable to FedAvg with perfect communications without requiring any instantaneous CSI thus alleviating the need for channel estimation and feedback.
These results highlight the advantage of NCAirFL in terms of trade-offs between learning performance and communications complexity.

\section{Conclusions}
This paper proposed NCAirFL, a CSI-free AirFL scheme based on unbiased non-coherent detection, which overcomes the challenge of signal alignment in frequency-selective fading channels without the knowledge of instantaneous CSI. 
By utilizing binary dithering and an error-compensation mechanism, NCAirFL achieves a convergence rate of $\mathcal{O}(1/\sqrt{T})$ after $T$ communication rounds for general non-convex and smooth objectives. 
Experiments confirmed that NCAirFL approaches the performance of FedAvg with a negligible gap in terms of test accuracy, offering a practical and efficient solution for FL in resource-constrained wireless networks.

\appendix \label{subsec:proof of theorem}
We apply the perturbed iterate analysis as in \cite{stich2018sparsified} to provide the convergence bound of NCAirFL. Define the maintained virtual sequence $\{\tilde{\boldsymbol \theta}^{(t)}\}_{t=0,\ldots,T-1}$ as follows:
$$
\tilde{\boldsymbol \theta}^{(0)}=\boldsymbol \theta^{(0)}, \ \tilde{\boldsymbol \theta}^{(t+1)} = \tilde{\boldsymbol \theta}^{(t)}- \frac{1}{rn}\sum_{i\in \mathcal{I}^{(t)}}\boldsymbol \Delta_i^{(t)} - \frac{\eta^{(t)}}{rn}\boldsymbol \phi^{(t)}\odot\boldsymbol e^{(t)},
$$
where $\boldsymbol e^{(t)}:=\boldsymbol r^{(t)}-(\Myfrac{1}{\eta^{(t)}})\sum_{i\in \mathcal{I}^{(t)}}\boldsymbol g_i^{(t)}$, and $\mathbb{E}[\boldsymbol e^{(t)}]=0$. \textcolor{black}{Hence, $\mathbb{E}[\boldsymbol \phi^{(t)}\odot\boldsymbol e^{(t)}] = \mathbb{E}_{\mv \phi^{(t)}}[\mathbb{E}[\boldsymbol \phi^{(t)}\odot\boldsymbol e^{(t)} \mid \mv \phi^{(t)}]]=0$.}
Following the global update of $\mv \theta^{(t)}$, 
$$\boldsymbol \theta^{(t+1)} \leftarrow \boldsymbol \theta^{(t)} - \frac{\eta^{(t)}}{rn}\boldsymbol \phi^{(t)}\odot \boldsymbol{r}^{(t)},$$ 
the relation between the real sequence $\{\mv \theta^{(t)}\}$ and the virtual sequence $\{\tilde{\mv \theta}^{(t)}\}$ is given by 
\begin{equation}
    \boldsymbol \theta^{(t)} - \tilde{\boldsymbol \theta}^{(t)} =\frac{1}{rn}\sum_{i=1}^{n}\boldsymbol m_i^{(t)}.
\end{equation}

Begin with $L$-smoothness, \textcolor{black}{conditioned on round $t$}, taking expectation over randomness from wireless channel, i.e., channel fading and AWGN noise, we obtain
\begin{align} \label{eq:L-smooth step}
    \mathbb{E} f(\tilde{\boldsymbol \theta}^{(t+1)}) \le & f(\tilde{\boldsymbol \theta}^{(t)}) - \mathbb{E}\left<\nabla f(\tilde{\boldsymbol \theta}^{(t)}), \frac{1}{rn}\sum_{i\in \mathcal{I}^{(t)}}\boldsymbol \Delta_i^{(t)}\right> \nonumber \\ 
    & \hspace{-0.2in} + \frac{L}{2}\mathbb{E}\left\| \frac{1}{rn}\sum_{i\in \mathcal{I}^{(t)}}\boldsymbol \Delta_i^{(t)}+\frac{\eta^{(t)}}{rn}\boldsymbol \phi^{(t)}\odot\boldsymbol e^{(t)} \right\|^2.
\end{align}
We first bound the inner-product term in \eqref{eq:L-smooth step}. Taking expectation \textcolor{black}{conditioned on round $t$} over the randomness of SGD, of device selection and of the sign-vector $\boldsymbol \phi^{(t)}$ obtains
\begin{align}
    & -\mathbb{E}\left<\nabla f(\tilde{\boldsymbol \theta}^{(t)}), \frac{1}{rn}\sum_{i \in \mathcal{I}^{(t)}} \boldsymbol \Delta_{i}^{(t)} \right> \nonumber \\ 
    & = -\mathbb{E}\left<\nabla f(\tilde{\boldsymbol \theta}^{(t)}), \frac{1}{n}\sum_{i=1}^{n}\sum_{q=0}^{Q-1}\eta^{(t)}{\nabla} f_i(\boldsymbol \theta_i^{(t,q)}) \right> \nonumber  \\ 
    & = -\eta^{(t)}Q \| \nabla f(\tilde{\boldsymbol \theta}^{(t)})\|^2 \nonumber \\ 
    & \, -\eta^{(t)}\mathbb{E}\left< \nabla f(\tilde{\boldsymbol \theta}^{(t)}),\frac{1}{n}\sum_{i=1}^{n}\sum_{q=0}^{Q-1}\left( {\nabla} f_i(\boldsymbol \theta_i^{(t,q)})-\nabla f_i(\tilde{\boldsymbol \theta}^{(t)}) \right) \right> \nonumber \\ 
    & \le -\frac{\eta^{(t)}Q}{2}\| \nabla f(\tilde{\boldsymbol \theta}^{(t)})\|^2 \nonumber \\ 
    & \quad + \frac{\eta^{(t)}}{2Q}\left\|\mathbb{E} \left[ \frac{1}{n}\sum_{i=1}^{n}\sum_{q=0}^{Q-1}\left( {\nabla} f_i(\boldsymbol \theta_i^{(t,q)}) -\nabla f_i(\tilde{\boldsymbol \theta}^{(t)}) \right) \right] \right\|^2. \nonumber 
\end{align}
Using Assumption \ref{assumption: L-smoothness}, \ref{assumption:bounded variance} and \ref{assumption:heterogeneity}, with the relation $\boldsymbol \theta^{(t)} - \tilde{\boldsymbol \theta}^{(t)} =\Myfrac{1}{(rn)}\sum_{i=1}^{n}\boldsymbol m_i^{(t)}$ (derived by the memory update rule and the definition of the maintained virtual sequence) and $\|\nabla f(\boldsymbol \theta^{(t)})\|^2 \le 2\|\nabla f(\boldsymbol \theta^{(t)})-\nabla f(\tilde{\boldsymbol \theta}^{(t)})\|^2 + 2\|\nabla f(\tilde{\boldsymbol \theta}^{(t)})\|^2$, combining the memory bound in \cite[Lemma 3]{basu19qsparse} (\textcolor{black}{$\mathbb{E}\|\boldsymbol m_i^{(t)}\|^2 \le \Myfrac{(4\eta^2(1-\lambda^2)Q^2G^2)}{\lambda^2}$}) derived by the contraction in \eqref{eq:contraction} with $\lambda=\min(p,1-p)$ and the multiple local update bound \cite[Lemma 2]{yang2021achieving} under the condition \textcolor{black}{$\eta^{(t)}=\eta \le \Myfrac{1}{(\sqrt{240}QL)}$,}
we have
{
\begin{multline} \label{eq:bound of inner-product term}
    -\mathbb{E}\left<\nabla f(\tilde{\boldsymbol \theta}^{(t)}), \frac{1}{rn}\sum_{i \in \mathcal{I}^{(t)}} \boldsymbol \Delta_{i}^{(t)} \right>  \le -\frac{\eta Q}{8}\| \nabla f(\boldsymbol \theta^{(t)})\|^2 \\
    +  5Q^2\eta^3 L^2 (\sigma_l^2+6Q\sigma_g^2) + \frac{6Q^3 L^2\eta^3(1-\lambda^2)}{r^2\lambda^2}G^2.
\end{multline}
}

We next bound the second-order term in \eqref{eq:L-smooth step}. Taking expectation \textcolor{black}{conditioned on round $t$} over the randomness of SGD, sign-vector $\boldsymbol \phi^{(t)}$, device selection, channel noise, and small-scale fading, we have
{
\begin{align} \label{eq:bound of second-order term}
& \frac{L}{2}\mathbb{E}\left\| \frac{1}{rn}\sum_{i\in \mathcal{I}^{(t)}}\boldsymbol \Delta_i^{(t)}+\frac{\eta}{rn}\boldsymbol \phi^{(t)}\odot \boldsymbol e^{(t)} \right\|^2 \nonumber \\ 
& = \frac{L}{2}\mathbb{E}\left\|\frac{1}{rn}\sum_{i\in \mathcal{I}^{(t)}}\boldsymbol \Delta_i^{(t)} \right\|^2 +\frac{\eta^2 L}{2r^2n^2}\mathbb{E}\left\|\boldsymbol \phi^{(t)}\odot\boldsymbol e^{(t)} \right\|^2 \nonumber \\
& \le \frac{L}{2} \eta^2 Q^2G^2+\frac{\eta^2 L}{2r^2n^2}\mathbb{E}\left\|\boldsymbol e^{(t)} \right\|^2,
\end{align}
}
where the first equality follows $\mathbb{E}[\boldsymbol e^{(t)}]=\mv 0$, and the second inequality follows $\mathbb{E}\|\mv \Delta_i^{(t)} \|^2 \le \eta^2 Q^2 G^2$ (derived by Jensen's inequality and Assumption~\ref{assumption:bounded variance}) and the error bound $\mathbb{E}\| \mv e^{(t)} \|^2 \le G_e^2$, where $G_e^2$ is defined in Theorem~\ref{theorem:convergence}. Due to page limits, the detailed proof of the explicit form of $G_e^2$ is omitted.

The final result is obtained by substituting \eqref{eq:bound of inner-product term} and \eqref{eq:bound of second-order term} into \eqref{eq:L-smooth step}, taking expectation over all randomness and taking telescope sum.

\addtolength{\topmargin}{-0.01in}
\bibliographystyle{IEEEtran}
\bibliography{DL_ref}

\end{document}